\documentstyle[12pt]{article}
\newcommand{\Aslash}{\not \!\! A}
\newcommand{\Dslash}{\not \!\! D}
\newcommand{\kslash}{\not \! k}
\newcommand{\dslash}{\not \! \partial}

\begin{document}

\makeatletter
\@addtoreset{equation}{section}
\def\theequation{\thesection.\arabic{equation}}
\makeatother

\begin{flushright}{January 2000\\UT-874}
\end{flushright}
\vskip 0.5 truecm

\begin{center}
{\large{\bf Regularization and Anomalies in Gauge 
Theory\footnote{Talk given at NATO Advanced Research Workshop 
``Lattice Fermions and Structure of the Vacuum'', October 5-9, 1999, 
at Dubna, Russia (To be published in the Proceedings)}
}}
\end{center}
\vskip .5 truecm
\centerline{\bf Kazuo Fujikawa}
\vskip .4 truecm
\centerline {\it Department of Physics,University of Tokyo}
\centerline {\it Bunkyo-ku,Tokyo 113,Japan}
\vskip 0.5 truecm

\begin{abstract}
Some of the basic issues related to the regularization and 
anomalies in gauge theory are reviewed, with particular 
emphasis on the recent development in lattice gauge theory.
The generalized Pauli-Villars regularization is discussed from
a view point of the covariant regularization of currents, and 
the construction of a regularized effective action in terms of 
covariant currents is compared with the lattice formulation of 
chiral Abelian theory.
\end{abstract}
\large 
\section{Introduction}
  The regularization is a fundamentally important issue of field 
theory with an infinite number of degrees of freedom. A closely 
related issue in field theory is quantum anomaly,
though the anomaly itself is perfectly finite. The anomaly is 
more closely related  to ``conditional convergence''in a loose 
sense, the boundary between divergence and convergence.
For this reason, the treatment of anomaly becomes rather subtle 
in a finite theory such as the lattice theory. 
In this talk, I briefly review some of the fundamental issues 
related to the regularization and amomalies from my own view 
point. I will discuss the continuum regularization as well as 
the lattice regularization, with particular emphasis on the 
recent exciting development in lattice gauge 
theory[1][2][3][4].     

\section{Brief review of continuum path integral}
We start with a brief summary of the continuum path integral
approach to chiral anomaly[5] and a  regularization 
which may be called ``mode cut-off''.

We study the QCD-type Euclidean path integral with $\Dslash\equiv
\gamma^{\mu}(\partial_{\mu}- igA^{a}_{\mu}T^{a}) =
\gamma^{\mu}(\partial_{\mu}- igA_{\mu})$,
\begin{equation}
\int{\cal D}\bar{\psi}{\cal D}\psi[{\cal D}A_{\mu}]
\exp[\int\bar{\psi}(i\Dslash-m)\psi d^{4}x + S_{YM}]
\end{equation}
where $\gamma^{\mu}$ matrices are anti-hermitian with 
$\gamma^{\mu}\gamma^{\nu} +
\gamma^{\nu}\gamma^{\mu} = 2g^{\mu\nu} = - 2\delta^{\mu\nu}$, 
and $\gamma_{5}
= - \gamma^{1} \gamma^{2} \gamma^{3} \gamma^{4}$ is hermitian. 
$S_{YM}$ stands for the Yang-Mills action and $[{\cal D}A_{\mu}]$ contains a suitable gauge fixing.   

To analyze the chiral Jacobian we expand the fermion variables
[5]
\begin{eqnarray}
\psi (x) &=& \sum_{n}a_{n}\varphi_{n}(x)\nonumber\\
\bar{\psi}(x) &=& \sum_{n}\bar{b}_{n}\varphi_{n}^{\dagger}(x)
\end{eqnarray}
in terms of the eigen-functions of hermitian $\Dslash$
\begin{eqnarray}
\Dslash\varphi_{n} (x) &=& \lambda_{n}\varphi_{n} (x)\nonumber\\
\int d^{4}x \varphi_{n}^{\dagger}(x)\varphi_{l} (x) &=& 
\delta_{n,l}
\end{eqnarray}
which diagonalize the fermionic action in (2.1).
The fermionic path integral measure is then written as 
\begin{equation}
{\cal D}\bar{\psi}{\cal D}\psi = \lim_{N\rightarrow \infty}
\prod_{n=1}^{N}
d\bar{b}_{n}da_{n}
\end{equation}
Under an infinitesimal  global chiral transformation
\begin{equation}
\delta\psi = i\alpha\gamma_{5}\psi, \ \ \ \delta\bar{\psi} = 
\bar{\psi}
i\alpha\gamma_{5}
\end{equation}
we  obtain the Jacobian factor
\begin{eqnarray}
J &=& \exp [-2i\alpha  \lim_{N\rightarrow\infty}\sum_{n=1}^{N}
\int d^{4}x \varphi_{n}^{\dagger}(x)\gamma_{5}\varphi_{n} (x)]
\nonumber\\
&=& \exp [-2i\alpha (n_{+} - n_{-})]
\end{eqnarray}
where $n_{\pm}$ stand for the number of eigenfunctions with vanishing
eigenvalues and  $\gamma_{5}\varphi_{n}= \pm \varphi_{n}$,
respectively, in (2.3). We here used the relation $\int d^{4}x 
\varphi_{n}^{\dagger}(x)\gamma_{5}\varphi_{n} (x)=0$ for 
$\lambda_{n}\neq 0$. The Atiyah-Singer index theorem 
$n_{+} - n_{-}= \nu$ with  
 Pontryagin index $\nu$, which was  confirmed for one-
instanton sector in $R^{4}$ space by Jackiw and 
Rebbi[6], shows that the chiral
Jacobian contains the correct information of chiral anomaly. 

To extract a local version of the index (i.e., anomaly), we start with the 
expression
\begin{eqnarray}
n_{+} - n_{-}&=& \lim_{N\rightarrow\infty}\sum_{n=1}^{N}
\int d^{4}x \varphi_{n}^{\dagger}(x)\gamma_{5}
f((\lambda_{n})^{2}/M^{2})\varphi_{n} (x)\nonumber\\
&=&\lim_{N\rightarrow\infty}\sum_{n=1}^{N}\int d^{4}x 
\varphi_{n}^{\dagger}(x)\gamma_{5}f(\Dslash^{2}/M^{2})
\varphi_{n} (x)\nonumber\\ 
&\equiv& Tr \gamma_{5}f(\Dslash{D}/M^{2})
\end{eqnarray}
for {\em any} smooth function $f(x)$ which rapidly goes to zero for $x=\infty$ with $f(0)=1$. Since 
$\gamma_{5}f(\Dslash^{2}/M^{2})$ is a well-regularized operator, we may now use the plane wave 
basis of fermionic variables to extract an explicit gauge field dependence,  and we  define  a 
local version of the index  as
\begin{eqnarray}
&&\lim_{M\rightarrow\infty}tr \gamma_{5}f(\Dslash^{2}/M^{2})
\nonumber\\
&\equiv&\lim_{M\rightarrow\infty}\sum_{n=1}^{\infty} 
\varphi_{n}^{\dagger}(x)\gamma_{5}f(\Dslash^{2}/M^{2})
\varphi_{n} (x)\nonumber\\
&=& \lim_{M\rightarrow\infty}tr \int \frac{d^{4}k}{(2\pi)^{4}}
e^{-ikx}\gamma_{5}f(\Dslash/M^{2})e^{ikx}\\
&=&\lim_{M\rightarrow\infty}tr \int \frac{d^{4}k}{(2\pi)^{4}}
\gamma_{5}f\{
(ik_{\mu}+ D_{\mu})^{2}/M^{2} - \frac{ig}{4}[\gamma^{\mu}, 
\gamma^{\nu}]F_{\mu\nu}/M^{2}\}\nonumber\\
&=&\lim_{M\rightarrow\infty}tr M^{4}\int 
\frac{d^{4}k}{(2\pi)^{4}}\gamma_{5}f\{
(ik_{\mu}+ D_{\mu}/M)^{2} - \frac{ig}{4}[\gamma^{\mu}, 
\gamma^{\nu}]F_{\mu\nu}/M^{2}\}\nonumber 
\end{eqnarray}
where the remaining trace stands for  Dirac and Yang-Mills indices. We also
used the relation
\begin{equation}
\Dslash^{2} = D_{\mu}D^{\mu} - \frac{ig}{4}[\gamma^{\mu}, 
\gamma^{\nu}]F_{\mu\nu}
\end{equation}
and the rescaling of the variable $k_{\mu}\rightarrow M k_{\mu}$. 

By noting  $tr \gamma_{5}= tr \gamma_{5}[\gamma^{\mu}, 
\gamma^{\nu}]=0$, the above expression ( after expansion in powers of $1/M$) is written as ( with $ 
\epsilon^{1234}=1$)
\begin{eqnarray}
\lim_{M\rightarrow\infty}tr \gamma_{5}f(\Dslash^{2}/M^{2})
&=&tr \gamma_{5}\frac{1}{2!}\{\frac{-ig}{4}[\gamma^{\mu}, 
\gamma^{\nu}]F_{\mu\nu}\}^{2} \int \frac{d^{4}k}{(2\pi)^{4}}
f^{\prime\prime}(-k_{\mu}k^{\mu})\nonumber\\
&&= \frac{g^{2}}{32\pi^{2}}tr \epsilon^{\mu\nu\alpha\beta}
F_{\mu\nu}F_{\alpha\beta}
\end{eqnarray}
where we used 
\begin{eqnarray}
\int \frac{d^{4}k}{(2\pi)^{4}}f^{\prime\prime}(-k_{\mu}k^{\mu})
&=&
\frac{1}{16\pi^{2}}\int_{0}^{\infty} f^{\prime\prime}(x)xdx
\nonumber\\
&=& \frac{1}{16\pi^{2}}
\end{eqnarray}
with $x= -k_{\mu}k^{\mu}>0$ in our metric. 
One can confirm that any finite interval $-L\leq k_{\mu} \leq L$ of momentum
variables in (2.8) {\em before} the rescaling $k_{\mu}
\rightarrow M k_{\mu}$
gives rise to a vanishing contribution to (2.10). In this sense, the short 
distance contribution determines the anomaly.

When one combines (2.7) and (2.10), one establishes the Atiyah-Singer index theorem (in $R^{4}$ 
space). 
We  note that the local version of the index (anomaly)  is valid for Abelian theory also.
The global index (2.7) as well as a local version of the index 
(2.10) are both independent of the regulator  $f(x)$  
provided[5] 
\begin{equation}
f(0) =1, \ \ \ f(\infty)=0,\ \ \ f^{\prime}(x)x|_{x=0}=
f^{\prime}(x)x|_{x=\infty}=0 
\end{equation}
If one chooses a smooth function $f(x)$ such that 
\begin{equation}
f(x)\simeq 1,\ \ \ 0\leq x\leq 1
\end{equation}
and $f(x)$ goes to $0$ very rapidly for $x> 1$, one has
\begin{equation}
\lim_{N\rightarrow\infty}\sum_{n=1}^{N}\varphi_{n}^{\dagger}(x)
\gamma_{5}f((\lambda_{n})^{2}/M^{2})\varphi_{n} (x)\simeq
\sum_{|\lambda_{n}|=0}^{|\lambda_{n}|\leq M}
\varphi_{n}^{\dagger}(x)\gamma_{5}\varphi_{n}(x)
\end{equation} 
The essence of the present regularization may thus be called 
gauge invariant ``mode cut-off'',following the terminology 
of Zinn-Justin.  

\section{Index theorem on the lattice}
We now come to the recent intersting development in lattice 
gauge theory. This development is based on the so-called
Ginsparg-Wilson relation[7]
\begin{equation}
\gamma_{5}D + D\gamma_{5}=aD\gamma_{5}D.
\end{equation}
where $a$ stands for the lattice spacing. If one defines the 
operator
\begin{equation}
\Gamma_{5}\equiv \gamma_{5}(1-\frac{1}{2}aD)
\end{equation}
which is hermitian, the above relation is written as 
\begin{equation}
\Gamma_{5}\gamma_{5}D +  \gamma_{5}D\Gamma_{5}=0.
\end{equation}
Namely, $\Gamma_{5}$ plays a role of $\gamma_{5}$ in continuum
theory. An explicit example of the operator $D$ which satisfies 
the Ginsparg-Wilson relation has been constructed by
Neuberger[1] and it is known as the overlap operator.

All the finite dimensional representations of the Ginsparg-Wilson
algebra (3.3)or the eigenstates $\phi_{n}$ of the hermitian 
$\gamma_{5}D$ 
\begin{equation}
\gamma_{5}D\phi_{n}=\lambda_{n}\phi_{n}
\end{equation}
 on a finite lattice are categorized into the following 3 classes:\\
(i)\ $n_{\pm}$ states,\\
\begin{equation}
\gamma_{5}D\phi_{n}=0, \ \ \gamma_{5}\phi_{n} = \pm \phi_{n},
\end{equation}
(ii)\ $N_{\pm}$ states($\Gamma_{5}\phi_{n}=0$), \\
\begin{equation}
\gamma_{5}D\phi_{n}= \pm \frac{2}{a}\phi_{n}, 
\ \ \gamma_{5}\phi_{n} = \pm \phi_{n},\ \ respectively,
\end{equation}
(iii) Remaining states with $0 < |\lambda_{n}| < 2/a$,
\begin{equation}
\gamma_{5}D\phi_{n}= \lambda_{n}\phi_{n}, 
\ \ \ \gamma_{5}D(\Gamma_{5}\phi_{n})
= - \lambda_{n}(\Gamma_{5}\phi_{n}), 
\end{equation}
and the sum rule $n_{+}+ N_{+} =  n_{-} + N_{-}$ holds[8]. 

All the $n_{\pm}$ and 
$N_{\pm}$ states are the eigenstates of $D$,  $D\phi_{n}=0$ and 
$D\phi_{n}= (2/
a) \phi_{n}$, respectively. If one denotes the number of states  
in (iii) 
by $2N_{0}$, the total number of states (dimension of the 
representation) $N$ is given by $N = 2(n_{+} + N_{+} +
N_{0})$, which is expected to be a constant independent of 
background gauge field configurations.

The index theorem on the lattice formulated by Hasenfratz, Laliena
and Niedermayer[2] is stated as the equality
\begin{equation}
Tr\Gamma_{5}=n_{+}-n_{-}=\nu
\end{equation}
in the continuum limit$a\rightarrow 0$.Here $n_{\pm}$ stand for 
the number of zero eigenvalue states in (3.5) with 
$\gamma_{5}\phi_{n}=\pm
\phi_{n}$, respectively, and $\nu$ stands for the Pontrygin index
or integrated form of chiral anomaly.The proof of this index
ralation proceeds as follows:\\
We first evaluate by using the above classification of states
\begin{eqnarray}
Tr\Gamma_{5}&=&\sum_{\lambda_{n}}\phi_{n}^{\dagger}\Gamma_{5}
\phi_{n}\nonumber\\
&=&\sum_{\lambda_{n}=0}\phi_{n}^{\dagger}\Gamma_{5}\phi_{n}
\nonumber\\
&=&\sum_{\lambda_{n}=0}\phi_{n}^{\dagger}\gamma_{5}\phi_{n}
=n_{+}-n_{-}
\end{eqnarray}
The explicit evaluation of $Tr\Gamma_{5}$ has been performed 
by various authors by perturbative calculation. The
result[2][9] confirms the relation
for $a\rightarrow 0$
\begin{equation}
Tr \gamma_{5}(1 - \frac{a}{2}D)(x)
=\int d^{4} \frac{g^{2}}{32\pi^{2}}tr 
\epsilon^{\mu\nu\alpha\beta}F_{\mu\nu}F_{\alpha\beta}. 
\end{equation}
The actual calculation is rather involved. 

We here present a 
somewhat simpler calculation[10], which is similar 
to the continuum calculation in Section 2. We start  with
\begin{equation}
Tr\{\gamma_{5}[1-\frac{1}{2}aD]
f(\frac{(\gamma_{5}D)^{2}}{M^{2}})\}
=n_{+} - n_{-}
\end{equation}
Namely, the index is not modified by any  regulator $f(x)$ with 
$f(0)=1$, as can be confirmed by using the basis in (3.5)-(3.7). 
The hermitian operator $\gamma_{5}D$ plays a priviledged role in 
the present analysis of the index theorem.
We then consider a local version of the index
\begin{equation}
tr\{\gamma_{5}[1-\frac{1}{2}aD]
f(\frac{(\gamma_{5}D)^{2}}{M^{2}})\}
\end{equation}
where trace stands for Dirac and Yang-Mills indices. 
A local version of the index is not sensitive to the precise boundary
condition , and  one may take the infinite volume limit 
$L=Na \rightarrow\infty$ in the above expression. 

We now examine the continuum limit $a\rightarrow 0$ of the above local expression 
(3.12)\footnote{This continuum limit corresponds to the so-called ``naive'' continuum limit in the 
context of lattice gauge theory.}. We first observe that the term
\begin{equation}
tr\{\frac{1}{2}a\gamma_{5}Df(\frac{(\gamma_{5}D)^{2}}{M^{2}})\}
\end{equation}
goes to zero in this limit. The large eigenvalues of $\gamma_{5}D$ are 
truncated at the value $\sim M$ by the regulator $f(x)$ which rapidly
goes to zero for large $x$. In other words, the global index of the
operator $Tr \frac{a}{2}\gamma_{5}D
f(\frac{(\gamma_{5}D)^{2}}{M^{2}})\sim O(aM)$.

We thus examine the small $a$ limit of 
\begin{equation}
tr\{\gamma_{5}f(\frac{(\gamma_{5}D)^{2}}{M^{2}})\}
\end{equation}
The operator appearing in this expression is well regularized by the function
$f(x)$ , and  we evaluate the above trace by using the plane wave basis to extract an explicit 
gauge field dependence.
We consider a square lattice where the momentum is defined in the Brillouin zone
\begin{equation}
-\frac{\pi}{2a}\leq k_{\mu} < \frac{3\pi}{2a}
\end{equation}
We assume that the operator $D$ is free of  species doubling; in other words, the operator $D$ 
blows up rapidly ($\sim \frac{1}{a}$) for small $a$ in the momentum region corresponding to species 
doublers. The contributions of doublers are  
eliminated by the regulator $f(x)$ in the above expression. We thus examine the above trace in the 
momentum range of the physical species
\begin{equation}
-\frac{\pi}{2a}\leq k_{\mu} < \frac{\pi}{2a}
\end{equation}

We now obtain the limiting $a\rightarrow 0$ expression
\begin{eqnarray}
&&\lim_{a\rightarrow 0}tr\{\gamma_{5}
f(\frac{(\gamma_{5}D)^{2}}{M^{2}})\}\nonumber\\
&=& \lim_{a\rightarrow 0}tr 
\int_{-\frac{\pi}{2a}}^{\frac{\pi}{2a}}\frac{d^{4}k}{(2\pi)^{4}}
e^{-ikx}\gamma_{5}f(\frac{(\gamma_{5} D)^{2}}{M^{2}})e^{ikx}
\nonumber\\
&=&\lim_{L\rightarrow\infty}\lim_{a\rightarrow 0}tr 
\int_{-L}^{L}\frac{d^{4}k}{(2\pi)^{4}}e^{-ikx}\gamma_{5}
f(\frac{(\gamma_{5} D)^{2}}{M^{2}})e^{ikx}\nonumber\\
&=&\lim_{L\rightarrow\infty}tr 
\int_{-L}^{L}\frac{d^{4}k}{(2\pi)^{4}}e^{-ikx}\gamma_{5}
f(\frac{(-i\gamma_{5}\Dslash)^{2}}{M^{2}})e^{ikx}\nonumber\\
&=&tr\{\gamma_{5}f(\frac{\Dslash^{2}}{M^{2}})\}
\end{eqnarray}
where  we first take the limit $a\rightarrow 0$ with fixed 
$k_{\mu}$ in 
$-L\leq k_{\mu} \leq L$, and then take the limit 
$L\rightarrow \infty$. This 
procedure is justified if the integral is well convergent[10].
 We also assumed that the operator $D$ satisfies  the following relation in the limit $a\rightarrow 
0$
\begin{eqnarray}
De^{ikx}g(x) &\rightarrow& e^{ikx}(\kslash -i\dslash -\Aslash)
g(x)\nonumber\\
&=& -i\Dslash(e^{ikx}g(x))
\end{eqnarray}
for any  {\em fixed} $k_{\mu}$, ($-\frac{\pi}{2a}< k_{\mu}< 
\frac{\pi}{2a}$), and a sufficiently smooth function $g(x)$. The function $g(x)$ corresponds to the 
gauge potential in our case, which in turn means that the gauge potential $A_{\mu}(x)$
is assumed to   vary very little  over the distances of the elementary lattice spacing. It is shown 
that an explicit example of $D$ 
given by Neuberger[1] satisfies the property (3.18) without species doublers.

Our final expression (3.17) in the limit $M\rightarrow\infty$  
thus reproduces the index theorem in the continuum formulation, 
(2.10), by using  the  quite general properties of the basic 
operator $D$ only: The basic relation (3.1) with hermitian 
$\gamma_{5}D$ and the continuum limit property (3.18) 
{\em without} species doubling in the limit $a\rightarrow 0$.

\subsection{Modified chiral transformation}
 Utilizing  the notion of the index on the lattice,  L\"{u}scher introduced a new kind of 
chiral transformation[3]
\begin{equation}
\delta\psi = i\alpha\gamma_{5}(1 -\frac{1}{2}aD)\psi, 
\ \ \ \delta\bar{\psi} = \bar{\psi}i\alpha (1 - \frac{1}{2}aD)
\gamma_{5} 
\end{equation}
with an infinitesimal constant parameter $\alpha$. This transformation leaves the action in 
\begin{equation}
\int {\cal D}\bar{\psi}{\cal D}\psi \exp [- \sum a^{4} 
\bar{\psi}D\psi ]
\end{equation}
invariant due to the property (3.1), and  gives rise to the chiral Jacobian 
factor
\begin{equation}
J = \exp \{ -2i\alpha Tr \gamma_{5}(1 -\frac{1}{2}aD)\}
\end{equation}
The index theorem (3.8) shows that this Jacobian factor indeed
carries the correct chiral anomaly.

As a generalization of the vector-like(QCD-type) theory discussed
so far,L\"{u}scher[11] showed that a chiral Abelian
gauge theory can be consistently defined on a lattice. In
particular, the anomaly in the fermion number current, which 
generally appears in chiral gauge theory, arises as a result 
of the non-vanishing index of a rectangular $m\times n$ matrix
$M$
\begin{equation}
dim\\ ker\\ M - dim\\ ker\\ M^{\dagger}= m-n
\end{equation}
Namely, the regularized Lagrangian for chiral fermion  is 
characterized
by a rectangular matrix instead of a naive square matrix. A 
further comment on the Abelian chiral theory on the lattice 
will be given later.

The lattice regularization of chiral non-Abelian gauge theory 
has not been formulated so far[12].   

\section{Generalized Pauli-Villars regularization}
A Lgrangian level regularization of chiral non-Abelian gauge 
theory in 
continuum has been formulated by Frolov and Slavnov[13],
and Narayanan and Neuberger[14]. This scheme is based 
on a generalization of the Pauli-Villars regularization. To 
regularize one chiral fermion, one needs to introduce an infinte
number of fermions and unphysical bosonic fermions.This 
regularization is applicable to anomaly-free gauge theory only.    

Instead of writing the regularized Lagrangian, we here discuss
the generalized Pauli-Villars regularization from a view point 
of the regularization of currents[15]. 
The essence of the generalized Pauli-Villars regularization is 
 summarized in terms of regularized currents as follows:\\
\begin{eqnarray}
\lefteqn{<\overline{\psi}(x)T^{a}\gamma^{\mu}
(\frac{1+\gamma_{5}}{2})\psi(x)>
_{PV}}\nonumber\\
&=&-\lim_{y{\rightarrow}x}\left\{\frac{1}{2}Tr\left[T^{a}
\gamma^{\mu}f(\not{\!\!D}^{2}
/\Lambda^{2})\frac{1}{i\not{\!\!D}}\delta(x-y)\right]\right.
\nonumber\\
& &\ \
\left.+\frac{1}{2}Tr\left[T^{a}\gamma^{\mu}\gamma_{5}
\frac{1}{i\not{\!\!D}}\delta
(x-y)\right]\right\}\nonumber\\
\lefteqn{<\overline{\psi}(x)\gamma^{\mu}(\frac{1+\gamma_{5}}{2})
\psi(x)>_{PV}}
\nonumber\\
&=&-\lim_{y{\rightarrow}x}\left\{\frac{1}{2}Tr\left[\gamma^{\mu}
f(\not{\!\!D}^{2}/\Lambda^{2})
\frac{1}{i\not{\!\!D}}\delta(x-y)\right]\right.\nonumber\\
& &\ \ \left.+\frac{1}{2}Tr\left[\gamma^{\mu}\gamma_{5}\frac{1}
{i\not{\!\!D}}
\delta(x-y)\right]\right\}\nonumber\\
\lefteqn{<\overline{\psi}(x)\gamma^{\mu}\gamma_{5}
(\frac{1+\gamma_{5}}{2})
\psi(x)>_{PV}}\nonumber\\
&=&-\lim_{y{\rightarrow}x}\left\{\frac{1}{2}Tr\left[
\gamma^{\mu}\gamma_{5}f(\not{\!\!D}^{2}/\Lambda^{2})
\frac{1}{i\not{\!\!D}}\delta(x-y)\right]\right.\nonumber\\
& &\ \ \left.+\frac{1}{2}Tr\left[\gamma^{\mu}\frac{1}
{i\not{\!\!D}}\delta(x-y)\right]\right\}.
\end{eqnarray}
where the regularization function is defined by
\begin{equation}
f(\Dslash^{2}/\Lambda^{2})=\sum_{n=-\infty}^{\infty}(-1)^{n}
\Dslash^{2}/[\Dslash^{2}+(n\Lambda)^{2}]=\frac{\pi(\Dslash/
\Lambda)}{sinh (\pi\Dslash/\Lambda)}
\end{equation}
Note that $f(x)$ satisfies all the properties in (2.12).
In the left-hand sides of these equations (4.1),the currents are 
defined in terms of the fields appearing in the original
chiral  Lagrangian, which one wants to regularize,while 
the right-hand
sides of these equations stand for the regularized expressions.
The axial-vector and vector $U(1)$ currents in terms of 
the chiral fermion fields in the original Lagrangian are 
identical if one note $\gamma_{5}^{2}=1$, but the regularized 
versions (i.e. the
last two equations in (4.1)) are different. In particular , the 
vector $U(1)$ current(i.e. , the second equation in (4.1)) is 
not completely regularized.
 We emphasize that all the one-loop diagrams are generated
 from the (partially) regularized currents in (4.1), as will be
discussed later in connection with the effective action ; in 
otherwords,(4.1) retains all the information of the generalized 
Pauli-Villars regularization[13][14]. It is 
interesting that this 
regularization is implemented in the Lagrangian level.

\subsection{Covariant regularization}
A closely related regularization of chiral currents is known as 
the 
covarianr regularization, which regularizes all the currents (and
consequently  all the one-loop fermionic diagrams) and reproduces 
the so-called covariant anomalies[15]. This covariant 
regularization is,however, not implemented in the Lagrangian 
level,in general. The currents in the covariant regularization 
are written as 
\begin{eqnarray}
\lefteqn{<\overline{\psi}(x)T^{a}\gamma^{\mu}
(\frac{1+\gamma_{5}}{2})\psi(x)>
_{cov}}\nonumber\\
&=&-\lim_{y{\rightarrow}x}\left\{ Tr\left[T^{a}\gamma^{\mu}
(\frac{1+\gamma_{5}}{2})f(\not{\!\!D}^{2}/\Lambda^{2})
\frac{1}{i\not{\!\!D}}\delta(x-y)\right]\right\}
\nonumber\\
\lefteqn{<\overline{\psi}(x)\gamma^{\mu}(\frac{1+\gamma_{5}}{2})
\psi(x)>_{cov}}
\nonumber\\
&=&-\lim_{y{\rightarrow}x}\left\{Tr\left[\gamma^{\mu}
(\frac{1+\gamma_{5}}{2})
f(\not{\!\!D}^{2}/\Lambda^{2})
\frac{1}{i\not{\!\!D}}\delta(x-y)\right]\right.\nonumber\\
\end{eqnarray} 
The difference of this regularization from the generalized 
Pauli-Villars regularization in (4.1) is that all the components 
( either 
vector or axial-vector) are well-regularized. All the fermionic
one-loop diagrams are thus regularized.
The price we have to pay for this is that this regularization is 
not implemented in the Lagrangain level.

The anomaly in the gauge current is given by
\begin{eqnarray}
D_{\mu}\lefteqn{<\overline{\psi}(x)T^{a}\gamma^{\mu}
(\frac{1+\gamma_{5}}{2})\psi(x)>_{cov}}&&\nonumber\\
&=&-D_{\mu}\sum_{n}\varphi_{n}(x)^{\dagger}T^{a}\gamma^{\mu}
(\frac{1+\gamma_{5}}{2})f(\lambda_{n}^{2}/\Lambda^{2})
\frac{1}{i\lambda_{n}}\varphi_{n}(x)\nonumber\\
&=&\sum_{n}(\Dslash\varphi_{n}(x))^{\dagger}T^{a}
(\frac{1+\gamma_{5}}
{2})f(\lambda_{n}^{2}/\Lambda^{2})
\frac{1}{i\lambda_{n}}\varphi_{n}(x)\nonumber\\
&&-\sum_{n}\varphi_{n}(x)^{\dagger}T^{a}(\frac{1-\gamma_{5}}{2})
f(\lambda_{n}^{2}/\Lambda^{2})
\frac{1}{i\lambda_{n}}\Dslash\varphi_{n}(x)\nonumber\\
&=&-i\sum_{n}\varphi_{n}(x)^{\dagger}T^{a}\gamma_{5}
f(\lambda_{n}^{2}/\Lambda^{2})\varphi_{n}(x)
\end{eqnarray}
where we used the eigenfunctions
\begin{equation}
\Dslash\varphi_{n}=\lambda_{n}\varphi_{n}
\end{equation}
We thus recover the Jacibian factor corresponding to the 
covariant anomaly. 

As for the fermion number anomaly, we have similarly 
\begin{eqnarray}
\partial_{\mu}\lefteqn{<\overline{\psi}(x)\gamma^{\mu}
(\frac{1+\gamma_{5}}{2})\psi(x)>_{cov}}&&\nonumber\\
&=&-i\sum_{n}\varphi_{n}(x)^{\dagger}\gamma_{5}
f(\lambda_{n}^{2}/\Lambda^{2})\varphi_{n}(x)\nonumber\\
&=&\partial_{\mu}<\overline{\psi}(x)\gamma^{\mu}\gamma_{5}
(\frac{1+\gamma_{5}}{2})\psi(x)>_{PV}
\end{eqnarray}
This shows that one can reproduce the correct fermion 
number anomaly by using the {\em axial} $U(1)$ current in (4.1) 
in the generalized Pauli-Villars regularization[16].

From this analysis, one can see that the generalized 
Pauli-Villars regularization is closely related to the covariant
regularization. Since the covariant regularization is applicable
to any chiral gauge theory, it is useful to decide if any theory 
is
anomalous or not.However, the covariant current as it stands
does not generate the integrable (or consistent) anomaly. This
issue is discussed in the next Section.

\section{Definition of effective action in terms of covariant
currents}
It is known that the effective action for the fermion is 
written in terms of the current. By using this fact, it has been
proposed by H.Banerjee, R.Banerjee and P.Mitra to write the 
regularized effective action in terms of the regularized 
covariant current[17]. As a simplest example, we 
discuss the Abelian chiral gauge theory defined by
\begin{eqnarray}
Z&=&\int{\cal D}\bar{\psi}{\cal D}\psi e^{\int d^{4}\bar{\psi}i
\Dslash(\frac{1+\gamma_{5}}{2})\psi}\nonumber\\
W&=&\ln Z = \ln det[i\Dslash(\frac{1+\gamma_{5}}{2})]\nonumber\\
\Dslash&=&\gamma^{\mu}(\partial_{\mu} -igA_{\mu})
\end{eqnarray}
We then obtain
\begin{equation}
\frac{\partial W}{\partial g}=\langle\int d^{4}x A_{\mu}(x)
\bar{\psi}(x)\gamma^{\mu}(\frac{1+\gamma_{5}}{2})\psi(x)\rangle
\end{equation}
The regularized effective action may be  defined in 
terms of the covariant current by 
\begin{equation}
W_{reg}\equiv\int_{0}^{g}\int d^{4}x A_{\mu}(x) 
 \langle\bar{\psi}(x)\gamma^{\mu}(\frac{1+\gamma_{5}}{2})\psi(x)
\rangle_{cov}   
\end{equation}
The consistent current is then derived from this regularized 
effective action as 
\begin{eqnarray}
j^{\mu}(x)_{cons}&\equiv&\frac{\delta}{\delta A_{\mu}(x)}
W_{reg}\nonumber\\
&=&\int_{0}^{g}dg j^{\mu}(x)_{cov}+\int_{0}^{g}dg\int dy 
A_{\nu}(y)\frac{\delta j^{\nu}(y)_{cov}}{\delta A_{\mu}(x)}
\\
&=&j^{\mu}(x)_{cov}-\int_{0}^{g}dgg
\frac{\partial}{\partial g}j^{\mu}(x)_{cov}+\int_{0}^{g}dg\int dy 
A_{\nu}(y)\frac{\delta j^{\nu}(y)_{cov}}{\delta A_{\mu}(x)}
\nonumber
\end{eqnarray}
Also
\begin{equation}
\frac{\partial}{\partial g}j^{\mu}(x)_{cov}=\int dy A_{\nu}(y)
\frac{\delta}{\delta(gA_{\nu}(y))}j^{\mu}(x)_{cov}
\end{equation}
as $j^{\mu}(x)_{cov}$ depends on $g$ only through the combination
$aA_{\nu}(y)$. We thus obtain
\begin{equation}
j^{\mu}(x)_{cons}=j^{\mu}(x)_{cov}+\int_{0}^{g}dg\int dy 
A_{\nu}(y)\{\frac{\delta j^{\nu}(y)_{cov}}{\delta A_{\mu}(x)}
-\frac{\delta j^{\mu}(x)_{cov}}{\delta A_{\nu}(y)}\}
\end{equation}
We note that by using (5.6) 
\begin{equation}
W_{reg}\equiv\int_{0}^{g}\int d^{4}x A_{\mu}(x)j^{\mu}(x)_{cov} 
=\int_{0}^{g}\int d^{4}x A_{\mu}(x)j^{\mu}(x)_{cons}    
\end{equation}
namely, the regularized effective action is independent of whether
the reguralized covariant current or regularized consistent 
current is used in its construction. All the naive properties are
reproduced by our definition of $W_{reg}$.

As for the chiral anomaly, we have(by noting that the Abelian 
covariant current is gauge invariant)
\begin{eqnarray}
W(A_{\mu}+\partial_{\mu}\omega)_{reg}&=&\int_{0}^{g}\int d^{4}x 
(A_{\mu}(x)+\partial_{\mu}\omega(x))j^{\mu}(x)_{cov}\nonumber\\
&=&W_{reg}-\int_{0}^{g}\int d^{4}x\omega(x) \partial_{\mu}
j^{\mu}(x)_{cov}
\end{eqnarray}
and if one lets the cut-off parameter $\Lambda\rightarrow\infty$ 
in the last covariant current, we generate the covariant anomaly
\begin{eqnarray}
W(A_{\mu}+\partial_{\mu}\omega)_{reg}&=&W_{reg} -
\frac{1}{16\pi^{2}}\int_{0}^{g}
\int d^{4}x\omega(x)F(gA_{mu})\tilde{F}(gA_{\mu})\nonumber\\
&=&W_{reg}-\frac{1}{3}\frac{1}{16\pi^{2}}\int d^{4}x\omega(x)
F(gA_{mu})\tilde{F}(gA_{\mu})
\end{eqnarray}
and we reproduce the consistent anomaly with the correct Bose
symmetrization factor$1/3$.It is known that this scheme 
works for the non-Abelian theory also[17].  

\subsection{Application to lattice gauge theory}
It has been pointed out by H. Suzuki[18] that the basic 
aspect of the
above construction of the regularized effective action in terms
of the covariant current works for the lattice theory also,
and one in fact ontains a formula closely related to the 
construction of the Abelian chiral theory given by 
L\"{u}scher[11]. 
 
The starting expression is
\begin{equation}
W=\int_{0}^{g}dg Tr\frac{\partial D}{\partial g}
(\frac{1+\hat{\gamma}_{5}}{2})D^{-1} 
\end{equation}
where $D$ stands for the lattice Dirac operator which satisfies
the Ginsparg-Wilson relation, and 
\begin{equation}
\hat{\gamma}_{5}\equiv\gamma_{5}(1-aD)
\end{equation}
with $(\hat{\gamma}_{5})^{2}=1$. A naive continuum limit 
 of (5.10) is 
\begin{equation}
W_{naive}=\int_{0}^{g}dg Tr\left[A_{\mu}\gamma^{\mu}
(\frac{1+\gamma_{5}}{2})\frac{1}{i\not{\!\!D}}\right]
\end{equation}
and $W$ is gauge invarinat in Abelian theory. Thus $W$ in (5.10) 
is a counter part of $W_{reg}$ in continuum theory.

It has been shown by Suzuki[18] that $W$ in (5.10)
for lattice theory
gives the first term of the consistent  lattice Abelian anomaly
\begin{equation}
\frac{1}{3}\frac{1}{16\pi^{2}}F\tilde{F}_{lattice} + 
\partial_{\mu}K^{\mu}
\end{equation}
where the second term is a ``lattice artifact'' found by 
L\"{u}scher[11]. Namely, $K^{\mu}$ is gauge invariant
and goes to $0$ in the naive continuum limit $a\rightarrow 0$.
An improvement of the above $W$ by using this $K^{\mu}$ has been
 shown [18] to be identical to the result in 
Ref.[11].

\section{Conclusion}
The remarkable development in lattice theory enriched our 
understanding of the regularization of fermions 
and the basic aspects of chiral symmetry and anomalies
in gauge theory. 
 It is interesting to see that the covariant current and 
consistent current play mutually complementary roles in these
constructions[19].
 
The interesting  notion of index on the 
lattice[2] deserves further investigation. The 
lattice formulation 
of chiral non-Abelian theory ( and eventually  supersymmetric 
theory) remains as a challenging problem[12].

\end{document}